\documentclass[12pt]{article}
\usepackage[utf8]{inputenc}
\usepackage{natbib}

\usepackage{url}
\usepackage{hyperref}

\usepackage{amsmath,array,graphicx}
\usepackage{textcomp}
\usepackage{gensymb}
\usepackage{float}
\usepackage{soul}
\usepackage{graphicx}
\usepackage{textgreek}
\usepackage{multicol, multirow}
\usepackage[many]{tcolorbox}
\usepackage{makecell}

\usepackage{booktabs}
\usepackage{xcolor}
\usepackage{enumitem}
\usepackage{siunitx}

\usepackage[labelfont=bf]{caption}
\usepackage[textfont=it]{caption}
\usepackage[margin=0.75in]{geometry}
\usepackage[superscript]{cite}
\usepackage{tabularx}
\usepackage{subcaption}
\definecolor{box_green}{RGB}{0,188,0}
\definecolor{box_purp}{RGB}{188,0,188}

\renewcommand{\hl}[1]{#1}

\newtcolorbox{promptquotebox}[1]{
  unbreakable,
  enhanced,
  colback=white,
  colframe=box_purp,
  arc=0pt,
  outer arc=0pt,
  title=#1,
  fonttitle=\bfseries\sffamily\large,
  colbacktitle=box_purp,
  attach boxed title to top left={},
  boxed title style={
    enhanced,
    skin=enhancedfirst jigsaw,
    arc=3pt,
    bottom=0pt,
    interior style={fill=box_purp}
  }
}

\title{Evaluating LLM-generated code for domain-specific languages: molecular dynamics with LAMMPS}

\author{Ethan W. Holbrook$^{a,*}$, Juan C. Verduzco$^{a,*}$, and Alejandro Strachan$^a$}

\date{
$^a$ School of Materials Engineering and Birck Nanotechnology Center \\
Purdue University, West Lafayette, Indiana, 47907 USA
\\
* Authors contributed equally to this work.
}

\begin{document}

\maketitle

\begin{abstract}

Large language models (LLMs) are changing the way researchers interact with code and data in scientific computing. While their ability to generate general-purpose code is well established, their effectiveness in producing scientifically valid \hl{scripts} for domain-specific languages (DSLs) remains largely unexplored. We propose an evaluation procedure that enables domain experts to assess the validity of LLM-generated input files for LAMMPS, a widely used molecular dynamics (MD) code, \hl{without requiring deep familiarity with its syntax. The evaluation procedure combines} a normalization step \hl{that produces} canonical \hl{input} files \hl{with} an extensible parser for syntax analysis\hl{, followed by a reduced-cost execution stage and accuracy checks that isolate common errors before running costly simulations. We apply the pipeline to eight state-of-the-art LLMs across three prompts of increasing complexity. The parser pass rate has improved from 74\% to 91\% over the past year, but scientific accuracy on coupled multi-step workflows remains limited. Across all 80 scripts evaluated on the most complex prompt, only one was fully correct as generated. We further package the automated stages as a reusable agentic skill that LLMs can invoke during script generation; in a small-scale demonstration, this skill helped two models produce five fully correct scripts out of six across the same three prompts, including the hardest one. The pipeline highlights both the limitations of current LLMs in generating scientific DSLs and a practical path toward integrating them into domain-specific computational ecosystems.}
\end{abstract}


\section*{Introduction}
\label{introduction_section}


Physics-based simulations are a staple in modern research and the development and availability of well-documented, verified, and efficient code, often tuned to take advantage of modern hardware, has firmly established computational science as a discipline. Examples include Quantum Espresso\cite{Giannozzi_2009} and VASP\cite{Kresse1996} for density functional theory calculations, LAMMPS\cite{thompson2022lammps} and NAMD\cite{Phillips2020} for molecular dynamics (MD), and MOOSE\cite{harbour2025moose} for finite elements. Running these tools requires knowledge of each software's ad hoc domain-specific language (DSL), the specialized interface through which numerical algorithms are translated into simplified commands describing physical models\cite{joel2024survey}. These DSLs are often syntactically complex. For example, even small deviations in command order or structure can invalidate a simulation. Errors often emerge when critical details are left implicit by relying on default values or through incomplete descriptions of scripts, an issue compounded by the scarce sharing of these input scripts. The associated semantic complexity, in contrast, reflects the need for a deep understanding of the underlying physics to construct meaningful and physically consistent simulation instructions. Together, these challenges hinder reproducibility, accessibility, and broader adoption. The pace of scientific discovery is slowed as a consequence. Bridging human intent with accurate and reproducible simulation workflows requires new strategies that reconcile natural language with the strict syntax of DSLs.

LAMMPS, the Large-scale Atomic/Molecular Massively Parallel Simulator, illustrates both the promise and the limitations of this paradigm. With its wide applicability and excellent performance \hl{across} multiple platforms, LAMMPS has unified a significant portion of the computational materials science community \hl{by exposing powerful MD capabilities\mbox{\cite{thompson2022lammps}} through an input language that shields users from the complexity of the underlying algorithms}, making simulations more accessible. Despite its impact, LAMMPS exemplifies the tension between flexibility and fragility in scientific DSLs: its ad hoc and semi-structured syntax mixes command dependencies and defaults in ways that make input files error-prone and difficult to validate. While researchers are generally able to articulate the simulations they wish to run, they may lack the expertise required to translate these descriptions into the proper syntax, and even experts can struggle to keep up with new developments. Recent advances in large language models (LLMs) offer a possible path forward, serving as interfaces between natural language descriptions and valid LAMMPS inputs. Yet, we lack procedures to evaluate, fix, and test LLM-generated LAMMPS scripts. 

In this work, we develop and test a procedure to evaluate LLM-generated input files for \hl{LAMMPS and demonstrate its utility across MD tasks of varying complexity. The pipeline is designed to enable domain experts to systematically identify and diagnose errors in LLM-generated scripts without requiring deep familiarity with the DSL.} The evaluation procedure starts with a normalization step that generates canonical inputs; this is followed by a static LAMMPS parser for syntax validation and early error detection and correction. Once an input file passes the parser, we use a series of simple tests to identify additional errors and evaluate the accuracy of the generated scripts. \hl{We apply this pipeline to a focused set of three prompts and eight widely used LLMs to illustrate its diagnostic capability across a range of simulation complexity levels.}

The evaluation procedure provides a framework for assessing how LLMs can be deployed in simulation-driven research, with an emphasis on reliability, interpretability, and reproducibility. Beyond LAMMPS, our approach points to a more general strategy for integrating LLMs into DSL-based scientific workflows, where bridging human intent and machine-readable inputs remains a critical challenge. \hl{In contrast to studies that focus on fine-tuning or domain-specific retraining, our work emphasizes the creation of independent evaluation and validation tools and their integration into the script generation loop, an approach that requires neither curated training data nor template scaffolding. To make this evaluation pipeline usable in practice, we additionally encapsulate its automated stages as a reusable agentic skill\mbox{\cite{zhang2025agentskills}}, integrated into Claude Code, that can be invoked by an LLM during script generation. This approach allows the pipeline to return structured, actionable feedback that agents can use to iteratively repair generated scripts and that humans can use to verify and validate their scientific accuracy.}

\section*{Related Work}
\label{relatedwork_section}

Beyond natural language tasks, LLMs have evolved into powerful problem solvers for both language and code generation. Models like GPT-5 \cite{openai2025gpt5systemcard}, Claude \cite{anthropic2025claudesystemcard}, and Gemini \cite{deepmind2025gemini_2_5_report} demonstrate state-of-the-art performance in general-purpose programming, whether they are generating functional code or tackling domain-specific problems. Researchers have validated these capabilities using standardized benchmarks going from simple coding exercises to realistic bug fixes \cite{xia2023automated}. This success largely stems from their ability to translate human requests into structured code or data formats, showcasing the versatility of LLMs in conventional programming contexts. 

Building on these foundational strengths, we are now seeing LLMs applied to concrete applications in science. Recent studies demonstrate that these models, when paired with specialized tools or agent-based frameworks, can extend beyond simple text generation to actually support real-world experiments and computational simulations. For instance, projects like CoScientist \cite{boiko2023autonomous} and ChemCrow \cite{m2024augmenting} have already demonstrated that LLMs can design and execute wet-lab protocols or plan chemical syntheses by integrating external tools. In computational fields, multi-agent frameworks like AtomAgents \cite{ghafarollahi2025automating} and DREAMS \cite{wang2025dreams} are proving that LLMs can plan, run, and analyze complex simulations in physics and materials science.

Despite these promising applications, a significant limitation remains: most of these systems remain narrow in scope. They rely heavily on tools and templates carefully curated by humans, rather than demonstrating the flexible, from-scratch generation we can get in general programming tasks \cite{joel2024survey,Gu2025}. In many cases, AI agents are tasked with preparing scripts for scientific software; however, they operate with expert-prepared packages and predefined templates \cite{mendible2025dynamate}. As a result, the task becomes less of one of autonomous generation and more one of structured assembly using curated components. This limitation is particularly evident when working with DSLs common in scientific software, which demand a much higher degree of transparency and correctness on their syntax than typical programming tasks \cite{joel2024survey,wang2023grammar}. Ultimately, this reliance on template-driven frameworks \hl{is less a design choice than a workaround for a deeper problem: the languages these frameworks wrap around were never built to be written or understood by anything other than expert researchers}.

Scientific DSLs present a distinct challenge for LLMs \hl{because they evolved without the supporting infrastructure that makes general-purpose languages tractable for code generation}. These specialized input languages, used in MD, quantum chemistry, or multi-scale simulators, are highly sensitive to syntax and remain underrepresented in the training of state-of-the-art models. \hl{They also lack the compilers, linters, and analysis tools that catch errors early in languages like Python and R}\cite{joel2024survey}. As a result, even minor mistakes in AI-generated scripts can crash simulations or produce results that look plausible but are scientifically invalid. Jacobs and Pollice \cite{jacobs2025developing} demonstrated this in their evaluation of LLMs on ORCA quantum chemistry input files, showing that while models can generate plausible inputs, they often failed to follow domain-specific conventions and syntax, especially without fine-tuning. Similarly, the FEABench project \cite{mudur2025feabench} found that LLMs struggled to generate reliable input scripts for finite-element analysis software such as COMSOL. Together, these findings underscore how far current systems are from producing trustworthy input scripts.

Among scientific DSLs, LAMMPS's input language provides a particularly instructive case\hl{, and recent work has explored multiple strategies for easing the burden it places on users.} The LAMMPS-GUI project \cite{lammps_tutorials_2025} has lowered the barrier to entry by providing a graphical interface with syntax highlighting, auto-completion, and inline help for configuring and running simulations. \hl{In service of these goals, AI-based approaches have attempted to translate natural-language descriptions into runnable scripts. MDAgent \mbox{\cite{shi2025fine}} is a fine-tuned LLM assistant that} can generate and review scripts for thermodynamic property calculations, \hl{illustrating} that domain-specific training \hl{can improve} output quality. \hl{Its follow-up, MDAgent2 \mbox{\cite{Shi2026}}, addresses two LAMMPS-specific challenges: the scarcity of domain-specific training data and the lack of established evaluation benchmarks for generated scripts. To address the latter, MDAgent2 introduces MD-EvalBench, a quiz-style benchmark that scores both structural and functional correctness of generated LAMMPS scripts.} Pushing automation further, DynaMate \cite{mendible2025dynamate} provides a modular multi-agent framework that manages the full molecular dynamics workflow from setup to analysis\hl{, coordinating} tools that use a Python package to convert high-level arguments into LAMMPS scripts. \hl{Across these AI-based systems, scripts are generated by capable models, but their correctness is handled in different ways: through self-evaluation, fine-tuning on curated data, or template-based scaffolding.}

\hl{What these approaches share is the absence of an independent check against the DSL itself. MDAgent and MDAgent2 rely on the AI models to evaluate their own outputs, while DynaMate sidesteps free-form generation by sending requests through its curated templates; neither case includes a structure-aware validation step between the model and the simulation. Combined with the absence of compiler and linter infrastructure, this leaves syntactic errors uncaught until the simulation runs. Self-evaluation is poorly suited to detect these errors, and template scaffolding avoids them only by restricting what can be generated in the first place. The evaluation stage carries a similar gap: MD-EvalBench reports relative scores across low-parameter models without a breakdown of where and why generation fails, which limits both diagnosis and improvement. Progress on AI-generated scripts for scientific DSLs therefore requires validation tooling that is independent of the generating model and integrated into the generation process itself.}

\hl{A practical mechanism for delivering this kind of validation has emerged in the form of \mbox{\textit{skills}\cite{zhang2025agentskills}}, which are structured, model-invocable tools that package logic into callable components an LLM-based agent can use during generation. Because the validation logic lives in the tool rather than in the model, this approach satisfies both properties identified above: it remains independent of the generating model while operating inside the generation loop. We adopt this approach in the present work, packaging our evaluation pipeline as a reusable skill that delivers per-stage diagnostic feedback to the model as it iterates on a script. More broadly, skill-based encapsulation offers a path for embedding domain-aware validation into agentic workflows for scientific DSLs where mature linter infrastructure does not yet exist.}

\section*{Methods}
\label{methods_section}

\subsection*{\hl{Pipeline}}
\begin{figure}[h!]
  \centering
  \includegraphics[width=.98\textwidth]{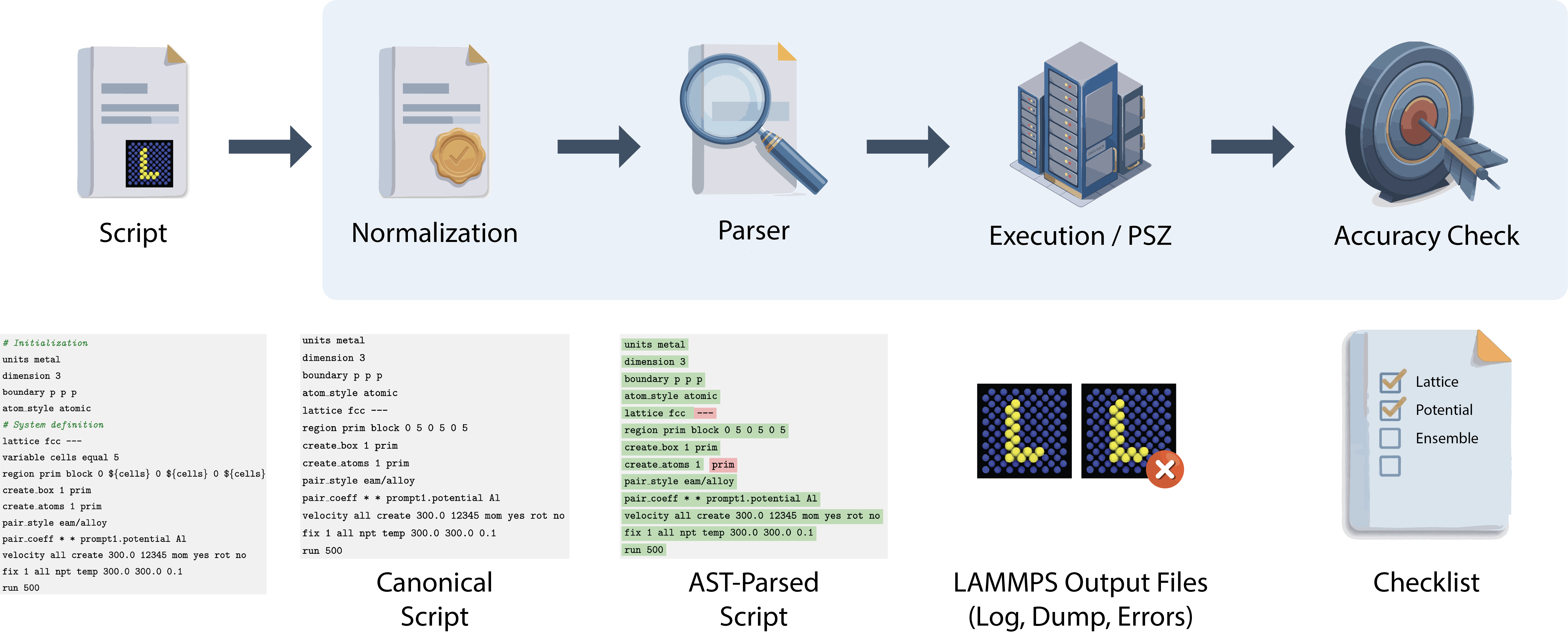}
  \caption{Workflow for evaluating LLM-generated LAMMPS input scripts. Each generated script passes through four stages: (1) normalization into a canonical form with comments removed, variables resolved, and loops expanded; (2) static syntax validation against a LAMMPS grammar, producing an abstract syntax tree (AST); (3) reduced-step execution, with pair\_style zero (PSZ) substitution applied on failure to isolate potential-specification errors; and (4) accuracy assessment against a prompt-specific checklist of physical criteria. Stages 1–3 run automatically; the final accuracy check is performed by a domain expert.}
  \label{fig:parser-workflow}
\end{figure}

To evaluate LLM-generated input files for LAMMPS, we developed and implemented a multi-stage procedure and evaluated it across tasks of varying complexity, as shown in Figure \ref{fig:parser-workflow}. The first stage involves script generation through each model's public API using default API parameters unless specified otherwise. All models received identical prompts and system instructions to ensure comparability. The system prompt, provided in the Supporting Information, was constructed using established prompt-design strategies such as explicit role specification, structured task instructions, contextual grounding, and chain-of-thought prompting \cite{wei2022chain, kojima2022large}. We note that this setup reflects guided rather than raw zero-shot performance, as would be expected in a realistic use case.

These generated input files are evaluated across three progressively complex prompts adapted from prior work \cite{verduzco2023gpt}. Prompt 1 represents a simple Aluminum single crystal equilibration at room temperature and pressure. Prompt 2 captures continuous heating of a bulk Ni single crystal from 300K to 2500K utilizing the isothermal-isobaric ensemble. Prompt 3, the most complex of the three, outlines a Niobium spall simulation with a projectile-gap-target impact setup. Together, these prompts span a modest range of complexity within the LAMMPS simulation space while focusing on simple single-metal systems rather than sampling the most complex multi-component, reactive, or molecular systems. \hl{These prompts were selected to span a representative range of single-element system complexities and simulation objectives.} The full text of the three evaluation prompts is shown in the Supporting Information. Post-processing of LAMMPS outputs is commonly performed using custom analysis workflows, so we excluded analysis-specific commands from evaluation to focus on simulation setup and execution logic. For each prompt–model combination, ten independent generations were produced (10-shot sampling) to capture stochastic variability in model outputs.

The language models used in this study are listed in Table \ref{tab:models}. All were accessed in their base configurations without fine-tuning or temperature adjustment. \hl{The effort levels were tested for the newer models from OpenAI, using a `high' effort for GPT-5.4 and the default `medium' effort level for GPT-5.5.}

\begin{table}[h!]
\centering
\caption{Language models evaluated in this study. All models were accessed via API in their base configuration without fine-tuning or temperature adjustment.}
\label{tab:models}
\begin{tabularx}{\textwidth}{llllX}
\toprule
\textbf{Provider} & \textbf{Name} & \textbf{Model ID} & \textbf{Release Date} & \textbf{Notes} \\
\midrule
OpenAI  & GPT-4o & gpt-4o-2024-08-06 & Aug 6, 2024 & \hl{Multimodal model with structured outputs} \\
OpenAI  & GPT-4.1 & gpt-4.1-2025-04-14 & Apr 14, 2025 & \hl{Coding and long-context (1M) model} \\
OpenAI  & \hl{o3} & o3-2025-04-16 & Apr 16, 2025 & \hl{First-tier reasoning model} \\
Anthropic & Opus 4 & claude-4-opus-20250514 & May 14, 2025 & \hl{Milestone agentic model} \\
OpenAI  & GPT-5 & gpt-5-2025-08-07 & Aug 7, 2025 & \hl{Unified reasoning model} \\
OpenAI & \hl{GPT-5.4(h)} & gpt-5.4-2026-03-05 & March 5, 2026 & \hl{Native computer-use agentic model} \\
Anthropic & \hl{Opus 4.7} & claude-opus-4-7 & April 16, 2026 & Flagship model \\
OpenAI & \hl{GPT-5.5(m)} & gpt-5.5-2026-04-23 & April 23, 2026 & Flagship model \\
\bottomrule
\end{tabularx}
\end{table}


The LLM-generated input files are evaluated using a multi-step approach. The first step is normalization, implemented within our `\textit{lammps-ast}' Python package and used to generate canonical input files for further evaluation. The normalization stage applies a sequence of preprocessing operations, including removal of comments and print statements, multi-line merging, and loop expansion followed by a set of routines for variable resolution, evaluation, and insertion. These steps remove extraneous content and normalize script structure, enabling a consistent and comparable representation for subsequent parsing and analysis. 

Handling variables is particularly challenging as both user-defined variables and built-in LAMMPS named parameters (such as thermodynamic properties, other predefined strings, \hl{or region arithmetic functions such as `count' which can only be evaluated at runtime}\cite{thompson2022lammps}) must be correctly interpreted. To enable reliable parsing, the normalizer evaluates variable expressions and replaces them with their corresponding numeric values, ensuring that arbitrary strings do not violate typing (integer or float) expectations for various arguments. Variable expressions that cannot be resolved will instead generate an error at the normalizer stage. This has the added benefit of ensuring identically valued scripts make up a unique tree with the parser, thereby enabling branch and leaf comparisons following the parsing stage. 

After normalization, the input scripts are processed by \textit{lammps-ast}, our custom parser for the LAMMPS input language built using the Lark Python package \cite{Lark2025}. The parser transforms the script and its LAMMPS commands, keywords, and argument lists into a typed abstract syntax tree (AST) representation. This process mirrors the compilation process for general-purpose languages, where the syntax tree serves as an intermediate representation for static analysis and semantic validation. By operating on a hierarchical tree representation rather than on linear text, the parser enables programmatic operations such as syntax inspection (for example, argument validation) and semantic consistency checking (for example, cross-referencing regions and variables). This approach allows malformed commands and incorrect arguments to be detected prior to MD simulation execution, so that syntactic inconsistencies can be identified and reported prior to simulation submission. \hl{The current version, available in the repository, incorporates a focused subset of LAMMPS commands spanning common simulation scenarios, including single- and multi-element equilibration, temperature ramps, energy minimization, mechanical deformation, and high-velocity impact setups.}

After parsing, all `run' statements in the scripts are adjusted to execute only 10 steps. This reduction limits the computational cost while still capturing most readily observable execution errors. For each script, the outcome of the LAMMPS run is recorded, and, in case of failure, the final line of the log file is captured as the error signal. We note that some simulations may complete the first 10 steps successfully even though they would fail at longer time scales and a simulation may run to completion and still be inaccurate. Running LAMMPS in this reduced mode tests script executability beyond basic syntax. This execution step also captures errors related to command ordering, which is not handled by the parser. One major source of failure during this execution step arises from inconsistencies during the setup of the interatomic potential used to describe interactions (\textit{pair\_style} parameters). Because these errors occur frequently and can obscure other issues, we introduced an additional step in which all pair\_style commands are replaced with \textit{``pair\_style zero"} (PSZ). This substitution separates errors related to syntax and variable handling from those tied to interatomic potential specification. After this modification, a second round of reduced-step execution is performed to continue assessing script executability and isolating sources of error. 

Finally, for scripts that executed without errors, either in their original form or after the PSZ substitution, we evaluated accuracy using prompt-specific checklists of required simulation conditions. For each prompt, we constructed quantitative criteria for the key physical parameters specified in the description by defining target values with an allowed tolerance. These criteria were organized around several fundamental aspects of the simulation setup. System definition included verification of the lattice parameters used to build the systems, boundary conditions, and the dimensions of the simulation cell. For the thermodynamic settings, we verified the ensemble used and the associated target temperatures and pressures. We also assessed the dynamics parameters, including the choice of timestep and the damping constants that regulate the system’s relaxation behavior. Finally, for the execution parameters, we examined run durations, heating or cooling rates, and the procedures used to initialize particle velocities. Each input file was evaluated against these criteria to determine whether the simulation would achieve the expected state or behavior. The criteria were iteratively refined to ensure coverage of all failure modes. This checklist-based assessment provides a systematic and reproducible measure of physical correctness that extends beyond mere syntactic or executability checks. 

\hl{Our evaluation pipeline operates automatically across the 
first three stages: normalization, static parsing, and reduced-step execution. Specifically, the normalizer processes raw LLM-generated scripts without human intervention, the parser performs syntax validation programmatically, and the LAMMPS execution stage carries out simulations without user input. The final accuracy assessment, which includes verifying that generated scripts satisfy the physical criteria specified in each prompt, was performed manually by domain experts. This human oversight step was intentional: scientific correctness cannot be fully captured by syntax checks or execution success alone, and expert judgment is required to determine whether a script would produce a physically meaningful simulation. This hybrid design reflects a realistic deployment scenario, where automated tooling handles structural validation while human experts retain responsibility for scientific validity.

Because the first three stages run without human intervention, they form a self-contained validation procedure that can be exposed to an LLM agent as a callable validation tool, allowing structured feedback to be returned during generation rather than only after a complete script has been produced. The next subsection describes this implementation.}

\subsection*{\hl{Agentic Skill}}

\hl{The evaluation pipeline's automated stages (normalization, static parsing, reduced-step execution) operate as standalone routines and can be packaged as a reusable skill that can be invoked within LLM-driven agentic workflows. This enables structured validation feedback to be delivered iteratively as the scripts are generated and refined, rather than only as a post-hoc assessment of a completed script. The skill returns actionable feedback that the agent can use as part of its decision-making, rather than reporting only pass/fail outcomes. We implement and demonstrate this design as a Claude Code \mbox{\cite{claudecode2025}} skill, but the same packaging pattern applies to any agentic framework that supports tool invocation.}

\hl{Upon invocation, the skill takes a LAMMPS input script as its only required input with optional configuration parameters such as the LAMMPS version and output directories. The skill walks through the pipeline starting from the normalization stage, processing the script through the same functions described in the pipeline. The normalized script is then sent to the static parser, which validates syntax and command structure against the LAMMPS grammar. For each identified error, the parser returns a structured error record containing the flagged line, its location, the offending token, and the associated error message. The skill also includes a reference guide that provides fix patterns for the most common failure modes identified in our pipeline experiments and instructs the agent to consult the official LAMMPS documentation for cases where verifying a specific command or unit convention is required. The skill additionally documents the limitations of the parser grammar to prevent valid commands from being misidentified as errors.}

\hl{Prior to the execution stage, the skill checks whether a LAMMPS executable is accessible through an environment variable or an explicit path argument. If identified, the skill proceeds with reduced-step execution to capture runtime errors related to command syntax and potential specification while minimizing computational cost. On execution failure, PSZ substitution is applied, replacing the pair style and pair coefficient commands with a non-operational potential, to determine whether the pair style specification is the isolated source of failure. If no executable is found, the skill reports this to the agent and terminates after the parser stage, exiting gracefully in environments where LAMMPS is not installed. The skill returns a structured JSON object containing per-stage status flags, parser error records (each specifying the flagged line, its position, and the unexpected token), and the normalized script for reference. Across repair iterations, these outputs are collected in a session log that can be rendered into a human-readable HTML or PDF report, documenting each repair round with per-stage outcomes, error details, and execution output.}

\section*{Results}
\label{results_section}

\begin{figure}[H]
  \centering
  \includegraphics[width=.98\textwidth]{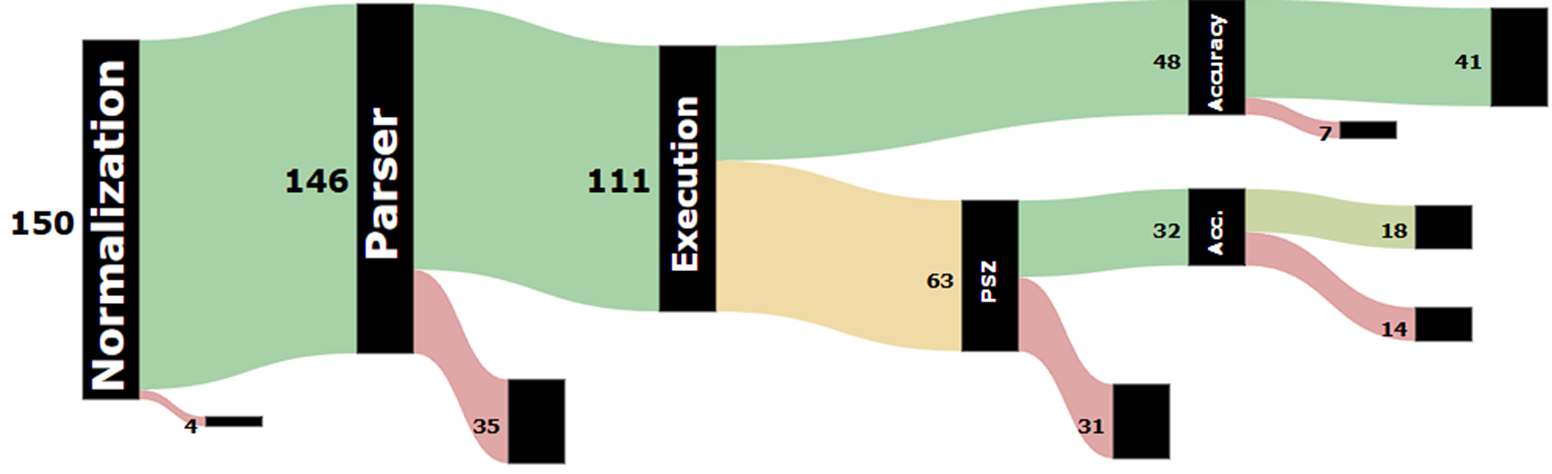}
  \caption{Sankey Diagram of the LAMMPS evaluation pipeline across all prompts and models \hl{examined with releases prior to 2026}. Flows track each generated script from normalization through parsing and execution to final accuracy checks.}
  \label{fig:Sankey}
\end{figure}

We evaluated a total of \hl{240} LLM-generated LAMMPS input scripts across all three prompts and \hl{eight} language models. \hl{We split the models by release date as performance is notably different for newer models.} These scripts were evaluated using the multi-stage assessment pipeline consisting of normalization, static parsing, reduced-step execution, and simulation accuracy checks. This pipeline allowed us not only to \hl{characterize error patterns and identify failures modes across different levels of simulation complexity, but also to demonstrate its utility as a diagnostic tool for LLM-generated DSL scripts}. Across \hl{all prompts and all models released prior to 2026}, approximately 74\% of the scripts passed the parser, with a smaller subset (32\%) executing without error on the first attempt, see Figure \ref{fig:Sankey}. Just over a quarter (27.3\%) of the scripts satisfied all the physical criteria specified in the prompts, with the majority corresponding to Prompt 1. This sharp drop in performance as the tasks become more complex illustrates the difficulty these LLMs face when generating inputs for scientific DSLs. Figure \ref{fig:Sankey} shows the number of scripts that pass each of the steps in our evaluation process, and a detailed breakdown for every prompt-model combination is provided in Table \ref{tab:model_performance_compact}.

\begin{table}[H]
\centering
\caption{LAMMPS evaluation pipeline outcomes by model and prompt. Execution and accuracy outcomes are separated into direct successes and additional successes obtained after PSZ substitution.}
\label{tab:model_performance_compact}
\small
\setlength{\tabcolsep}{4.5pt}
\renewcommand{\arraystretch}{1.12}
\begin{tabular}{llrrrrrr}
\toprule
\multirow{2}{*}{\textbf{Model}} &
\multirow{2}{*}{\textbf{Prompt}} &
\multirow{2}{*}{\textbf{Normalization}} &
\multirow{2}{*}{\textbf{Parsing}} &
\multicolumn{2}{c}{\textbf{Execution}} &
\multicolumn{2}{c}{\textbf{Accuracy}} \\
\cmidrule(lr){5-6} \cmidrule(lr){7-8}
 &  &  &  & \textbf{Direct} & \textbf{PSZ} & \textbf{Direct} & \textbf{PSZ} \\
\midrule

\textbf{gpt-4o} & TOTAL & 30 & 23 & 8  & 8  & 7  & 3 \\
                & P1    & 10 & 10 & 7  & 3  & 7  & 3 \\
                & P2    & 10 & 7  & 1  & 5  & 0  & 0 \\
                & P3    & 10 & 6  & 0  & 0  & 0  & 0 \\
\midrule

\textbf{gpt-4.1} & TOTAL & 30 & 24 & 8  & 13 & 7  & 4 \\
                 & P1    & 10 & 10 & 7  & 3  & 7  & 3 \\
                 & P2    & 10 & 9  & 1  & 8  & 0  & 1 \\
                 & P3    & 10 & 5  & 0  & 2  & 0  & 0 \\
\midrule

\textbf{o3} & TOTAL & 29 & 17 & 10 & 0  & 8  & 0 \\
                & P1    & 10 & 6  & 4  & 0  & 4  & 0 \\
                & P2    & 9  & 8  & 5  & 0  & 4  & 0 \\
                & P3    & 10 & 3  & 1  & 0  & 0  & 0 \\
\midrule

\textbf{claude-4-opus} & TOTAL & 30 & 29 & 10 & 10 & 9  & 8 \\
                       & P1    & 10 & 10 & 10 & 0  & 9  & 0 \\
                       & P2    & 10 & 10 & 0  & 10 & 0  & 8 \\
                       & P3    & 10 & 9  & 0  & 0  & 0  & 0 \\
\midrule

\textbf{gpt-5} & TOTAL & 27 & 18 & 12 & 3  & 10 & 3 \\
               & P1    & 10 & 10 & 8  & 0  & 6  & 0 \\
               & P2    & 7  & 7  & 3  & 3  & 3  & 3 \\
               & P3    & 10 & 1  & 1  & 0  & 1  & 0 \\
\midrule

\textbf{gpt-5.4} & TOTAL & 28 & 24 & 8  & 0  & 7*  & 0* \\
                 & P1    & 10 & 10 & 5  & 0  & 5*  & 0* \\
                 & P2    & 10 & 9  & 2  & 0  & 2*  & 0* \\
                 & P3    & 8  & 5  & 1  & 0  & 0*  & 0* \\
\midrule

\textbf{claude-4-7-opus} & TOTAL & 30 & 30 & 24 & 0  & 20* & 0* \\
                         & P1    & 10 & 10 & 10 & 0  & 10* & 0* \\
                         & P2    & 10 & 10 & 10 & 0  & 10* & 0* \\
                         & P3    & 10 & 10 & 4  & 0  & 0*  & 0* \\
\midrule

\textbf{gpt-5.5} & TOTAL & 29 & 28 & 18 & 5  & 14* & 5* \\
                 & P1    & 10 & 10 & 10 & 0  & 10* & 0* \\
                 & P2    & 9  & 9  & 4  & 5  & 4*  & 5* \\
                 & P3    & 10 & 9  & 4  & 0  & 0*  & 0* \\

\bottomrule
\end{tabular}
\end{table}

\begin{figure}[H]
  \centering
  \includegraphics[width=.98\textwidth]{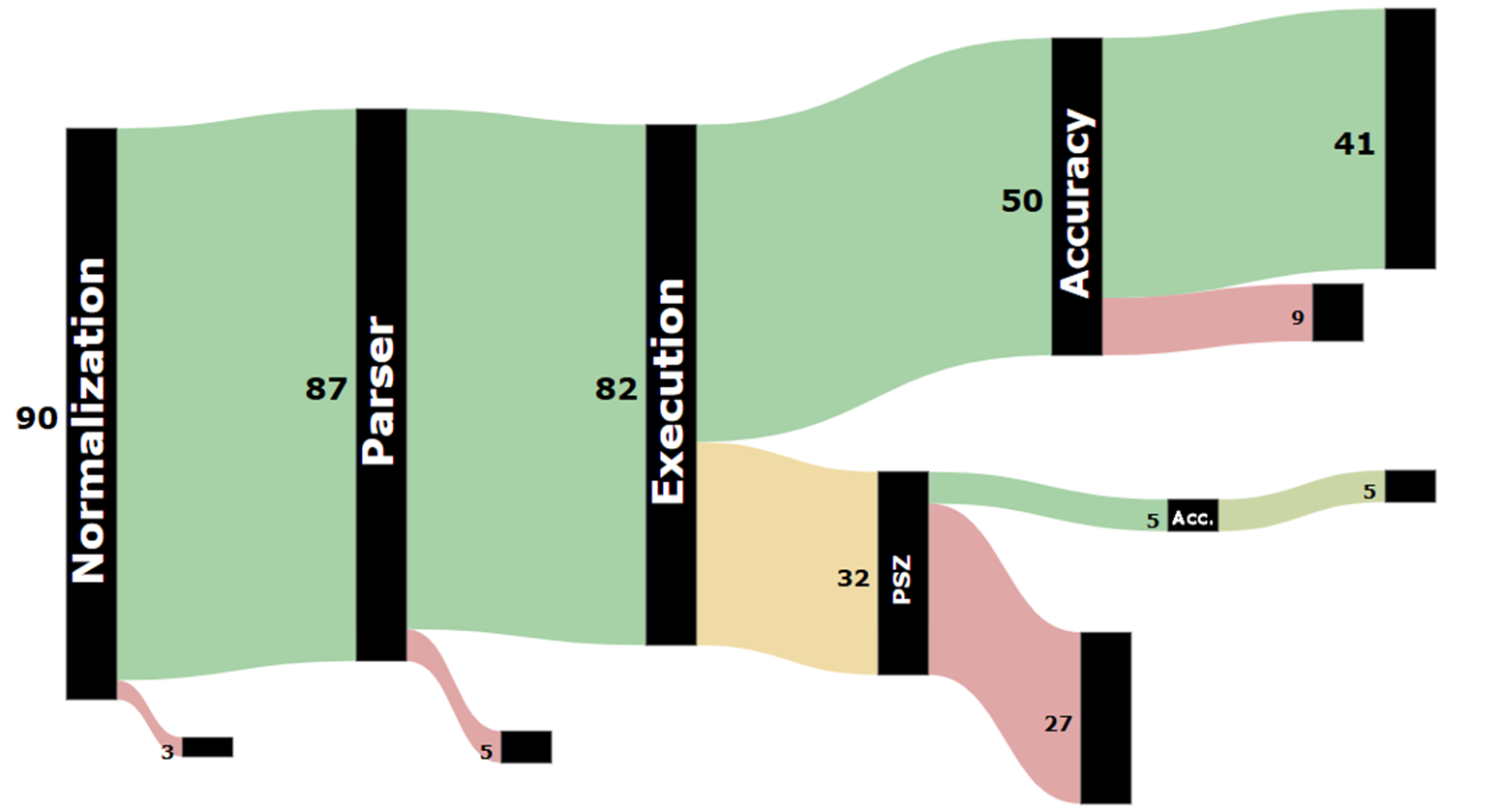}
  \caption{Sankey Diagram of the LAMMPS evaluation pipeline across all prompts and the newest models. Flows track each generated script from normalization through parsing and execution to final accuracy checks.}
  \label{fig:Sankey2}
\end{figure}

\hl{The newer models show substantially improved structural reliability: 91\% passed the parser and 55\% executed successfully on the first attempt. A representative breakdown for GPT-5.4, GPT-5.5, and Claude 4.7 Opus is shown in Figure \mbox{\ref{fig:Sankey2}}. Applying the same automated custom-built accuracy criteria used for the older cohort, we record a 46\% accuracy, though these results have not yet been verified by domain experts and are marked with an asterisk in Table \mbox{\ref{tab:model_performance_compact}}; the older models, in contrast, were hand-verified because automated accuracy checking cannot reliably distinguish the many ways a LAMMPS script can be correct or incorrect. Most of the gains concentrate in Prompts 1 and 2, while Prompt 3 remains difficult across all models. Going forward, we expect that monitoring simulation outputs, rather than checking parameters against a fixed checklist, will provide a more comprehensive measure of accuracy. Even so, most generated scripts contain only a few errors, and serve as reasonable starting points, reinforcing the need for a rigorous evaluation procedure for LLM-generated DSL scripts.}

To compare model behavior across the full LAMMPS code-generation pipeline, we evaluate each model using three aggregate metrics: \textit{parser pass rate} (the fraction passing normalization and static parsing), \textit{execution success rate} (the fraction completing a full LAMMPS run, including scripts that required PSZ correction), and \textit{one-shot accuracy} (the fraction fully correct without modification). Execution success indicates that a LAMMPS run completed but does not guarantee the simulation matches the desired outcome described in the prompt. Aggregate results are reported in Table \ref{tab:model_metrics}. \hl{Inference costs and token utilization for these runs are reported in the Supporting Information, Table \mbox{\ref{tab:costs}}.}

\hl{Among the older cohort, Claude Opus 4 leads on parser reliability (29/30) and execution success, with most of its Prompt 2 runs completing only after a PSZ correction. GPT-4.1 and GPT-4o perform well on the simplest prompt but degrade sharply on more complex workflows; their most common errors involve incorrect \textit{pair\_style} usage, malformed math expressions, and invalid region or group definitions. Both still maintain high parser pass rates and reasonable execution success, with many runs completing after only a minor PSZ fix. OpenAI's o3 and GPT-5 models show reduced robustness across the pipeline: GPT-5 produces most normalizer failures due to malformed variable expressions, while o3 has the weakest parser and second weakest execution rates overall, even though it does better on Prompt 2 than some other models. Common issues for both models include hallucinated command arguments, invalid variable expressions, and structural inconsistencies that prevent scripts from parsing or running. Notably, GPT-5 was the only model to produce a fully accurate Prompt 3 script and achieved the highest one-shot accuracy rate of the hand-verified cohort, while o3 came within a single error of a correct Prompt 3 script.} 

\hl{Among the newer models, Claude Opus 4.7 shows the strongest overall performance. It is the only model whose scripts passed every stage of the pipeline on Prompts 1 and 2, and it also achieved one of the highest execution success rates on Prompt 3. Its failures concentrate on Prompt 3, where incorrect region definitions and boundary-condition errors cause runs to simulate the wrong system or stop after only a few steps. This largely mirrors the performance of Claude Opus 4 with some improvement. GPT-5.5 performs well across the board, achieving the second highest execution rate, one script behind Claude Opus 4.7 when PSZ corrections are included. GPT-5.4 ranks below many of the older models; its scripts were plagued by an issue where run\_style commands were placed out-of-order before the simulation box definition. GPT-5.4 had the worst execution rate of any model despite parser pass rates just above the bottom of the set. We tested GPT-5.5 at the default `medium' effort and GPT-5.4 at `high' effort, which may partly explain GPT-5.4's weak showing, but the underlying cause is unclear. None of the newer models matched GPT-5's fully accurate Prompt 3 script in the hand-verified set, though their execution rates meet or exceed its performance.}

\begin{table}[H]
\centering
\caption{{Aggregate model performance across all three prompts. For each model, we report the number of scripts that 1) pass parsing, 2) execute successfully (with PSZ change made when necessary), and 3) are fully accurate as-generated.}}
\label{tab:model_metrics}
\begin{tabular}{lrrr}
\toprule
Model &  Parser pass rate &  Execution success rate & One-shot Accuracy \\


\midrule

GPT-4o          & 23/30 (77\%) & 14/30 (47\%) & 7/30 (23\%)\\
GPT-4.1         & 24/30 (80\%) & 21/30 (70\%) & 7/30 (23\%)\\
o3          & 17/30 (57\%) & 10/30 (33\%) & 8/30 (27\%)\\
Claude Opus 4   & 29/30 (97\%) & 20/30 (67\%) & 9/30 (30\%) \\
GPT-5           & 18/30 (60\%) & 15/30 (50\%) & 10/30 (33\%)\\
GPT-5.4          & 24/30 (80\%)   & 8/30 (27\%)  & 7/30* (23\%) \\
Claude Opus 4.7  & 30/30 (100\%)  & 24/30 (80\%) & 20/30* (66\%) \\
GPT-5.5          & 28/30 (93\%)   & 23/30 (77\%) & 14/30* (47\%) \\

\bottomrule
\end{tabular}
\end{table}




Prompt-level performance reveals clear differences in how well current language models handle increasingly complex LAMMPS workflows, as summarized in Table \ref{tab:prompt_metrics}. Prompt 1 consisted of a relatively simple isothermal/isobaric (NPT ensemble) equilibration of an Al supercell using conventional MD settings and a standard EAM potential. It was the least complex task in the study and produced consistently strong performance, with high parser pass rates \hl{(76/80)}, reliable execution (including PSZ corrections) \hl{(67/80)}, and the majority of all one-shot correct scripts in our analysis \hl{(58/80)}. Most models produced valid MD setups with only minor variations in quality. The most common error observed across all Prompt 1 scripts was a mismatch in the pair\_style definitions, where several models incorrectly specified \textit{eam} rather than \textit{eam/alloy}, a more common choice based on our experience. Despite this recurring issue, Prompt 1 remained the task where models showed the highest stability and correctness.

\begin{table}[h!]
\centering
\caption{Aggregate prompt performance across all models. For each prompt, we report the number of scripts that 1) pass parsing, 2) execute successfully (with PSZ change made when necessary), and 3) are fully accurate as-generated.}
\label{tab:prompt_metrics}
\begin{tabular}{lrrr}
\toprule
Prompt &  Parser pass rate &  Execution success rate & One-shot Accuracy \\


\midrule

Prompt 1  & 76/80 (95\%) & 67/80 (84\%) &  58/80 (72\%)\\
Prompt 2  & 69/80 (86\%) & 57/80 (71\%) &  23/80 (29\%)\\
Prompt 3  & 48/80 (60\%) & 13/80 (16\%)  &   1/80 (1\%) \\

\bottomrule
\end{tabular}
\end{table} 

Prompt 2 asked the models to simulate Ni melting by applying a continuous temperature ramp from 300 K to 2500 K under NPT conditions using a Mishin \cite{Mishin1999} EAM potential. This task introduced more complexity than the previous one, and model performance declined accordingly. Although the parser pass rate remained high \hl{(69/80)}, the execution success rate (including PSZ corrections) dropped moderately \hl{(57/80)}, and only a few scripts \hl{(23/80)} achieved one-shot correctness. The most common issues again involved problems in the pair\_style definitions \hl{(29/80)}, only around half of which could be considered accurate when ignoring pair\_style issues because the remaining cases contained additional inaccuracies. Other frequent errors included incorrect heating-rate specification and misconfigured thermostat and barostat damping constants. Prompt 2 revealed that even when syntax was largely correct, models still struggled to reliably encode the coupled temperature-ramp and ensemble specifications required for a melting simulation. The basic mathematical operations required to determine the rates, other simulation parameters, and unit conversions remain challenging for LLMs to handle.

\begin{figure}[H]
  \centering
  \includegraphics[width=.98\textwidth]{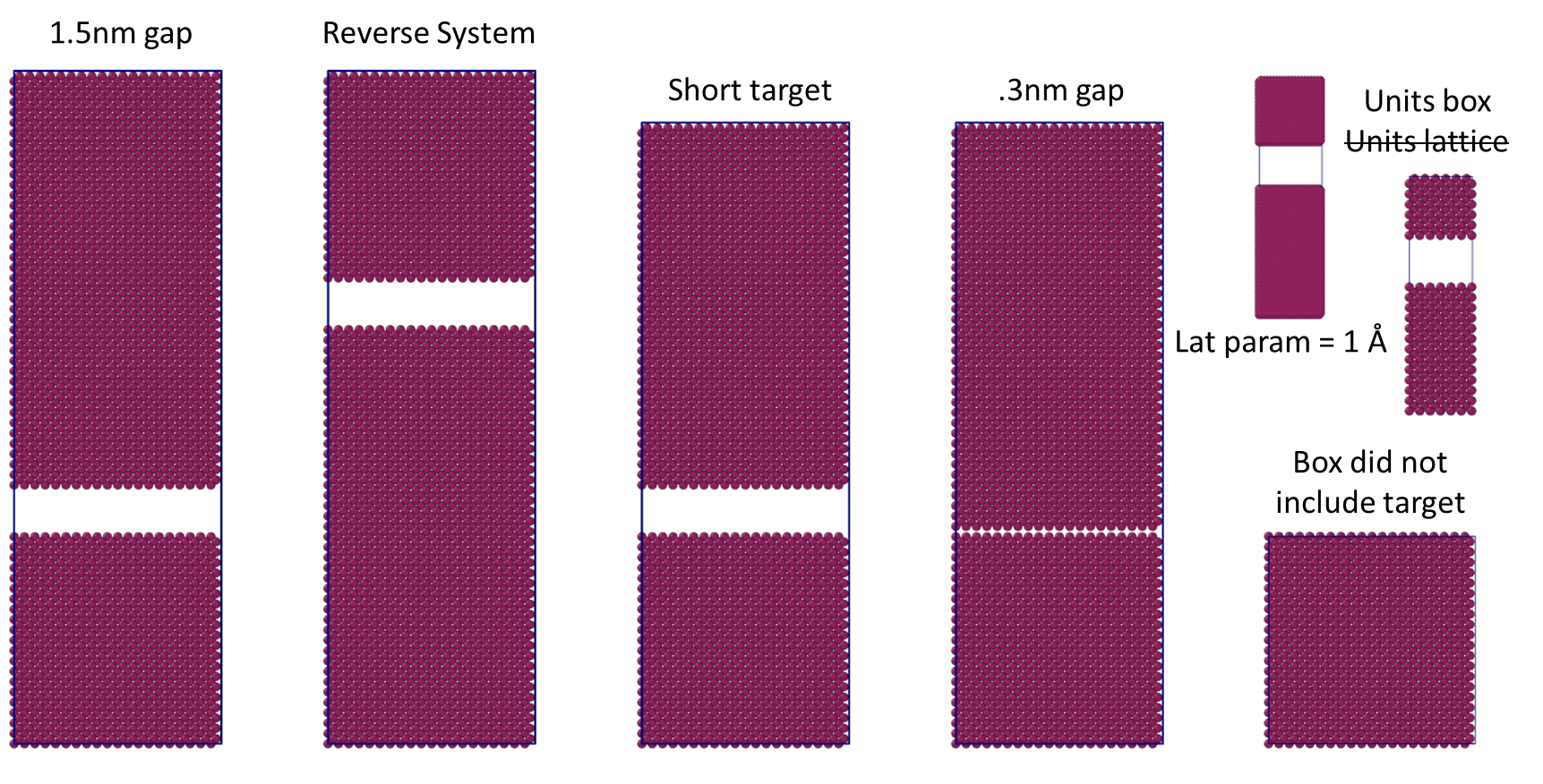}
  \caption{Representative failure modes observed for Prompt 3. Errors include incorrect gap specification, reversed system orientation, incorrect target region specification, improper unit handling (box vs lattice units), and omission of the target from the simulation box.}
  \label{fig:sim-errors}
\end{figure}

Prompt 3 required the models to set up a high-velocity projectile–target impact simulation in Nb at 2 km/s, making it the most complex workflow in our study. Model performance dropped sharply under this task. About half of the scripts passed parsing \hl{(48/80)}, only \hl{thirteen} executed (two with the PSZ correction), and only one script achieved one-shot correctness. The dominant failure modes again involved pair\_style definitions and errors in region and group definitions. System-setup inconsistencies produced widely varying geometries most commonly from miscalculations of the projectile–target sizes and the 1.5 nm gap, as shown in Figure \ref{fig:sim-errors}. Overall, Prompt 3 shows that models struggle to coordinate the multiple structural and dynamical specifications required for an impact simulation of this complexity; nevertheless, the single correct script indicates that such workflows are achievable, though not in a reliable or consistent manner. \hl{Models are showing improvement in parsing and execution over time and accuracy is expected to follow. The agentic workflow shows that fully accurate scripts for even prompt 3 are becoming a reasonable expectation.}

\hl{To demonstrate the applicability of the evaluation pipeline as a skill, we ran a single agentic generation for each of the three prompts using two models, Claude Sonnet 4.6 and Claude Opus 4.7, deployed through Claude Code. The skill was loaded at the start of each session so the model could invoke it as part of its generation process. These agentic workflows were provided with an initial prompt that mirrors our pipeline experiments; the conversations did not iterate after the first run of the skill to reduce variability introduced by natural language follow-up prompts. We note that, at the time of publishing, Opus 4.7 does not provide to its users the internal reasoning steps. All six resulting scripts were manually reviewed by a domain expert against the same accuracy criteria used for the one-shot evaluation, given the small sample size. Inference costs and token utilization for these runs are reported in the Supporting Information, Table \mbox{\ref{tab:agentic-tokens}}, and the full transcript of each conversation is included in the GitHub repository.}

\hl{Of the six scripts produced, five were judged fully accurate. Opus 4.7 generated correct scripts for all three prompts; Sonnet 4.6 succeeded on Prompts 1 and 2 but failed on Prompt 3. The failure in the script generated by Sonnet 4.6 was in setting the box up with an extra lattice spacing multiplication factor making the system 3.3 times larger in each direction; however, this would only be discoverable by the important validation and verification step performed by humans after the evaluation pipeline, either through examining the thermodynamic quantities or visualizing the system. Additional noteworthy choices were the velocity being initialized at 600 K for an equilibration at 300 K in Prompt 1 for Opus 4.7, and a timestep of 0.002 used in combination with the proper number of steps for a correct heating rate in Prompt 2 for Sonnet 4.6. Both models invoked the skill but performed little iterative repair within the loop, as their initial scripts already passed the normalization, parser, and reduced-step execution stages with few errors. In this regime, the skill functioned more as a pre-execution validation layer than as an active repair mechanism.}

\hl{Several qualitative patterns emerged in the agents' behavior. Sonnet 4.6's reasoning traces showed extended deliberation about LAMMPS syntax before invoking the skill on a candidate script, including discussion of command structure and ordering that suggests prior exposure to the DSL during training. Notably, neither model used its available web search tool to consult the LAMMPS documentation, instead relying on internal knowledge supplemented by the skill's reference guide, which encodes remediation patterns derived from the failure modes identified in our pipeline experiments. This pattern of internal reasoning paired with skill invocation, without external lookup, describes how the agents actually used the validation tooling in practice.}

\section*{Discussion}

\hl{Our evaluation reveals a clear pattern:} current LLMs can generate syntactically valid LAMMPS input scripts and capture key elements of standard MD workflows. However, their performance degrades sharply as simulation complexity increases and tasks require a larger number of commands, more intricate mathematical specifications, and increasingly complex geometries. \hl{Across all models, accuracy fell sharply between Prompt 1 (basic equilibration) and Prompt 2 (temperature ramp), and dropped further still on Prompt 3, which combined multi-region geometry, non-equilibrium dynamics, and group-specific operations with unit conversions and arithmetic that proved consistently challenging.} \hl{This performance drop indicates} that present-day LLMs, when used in isolation, are not yet able to execute the multi-constraint reasoning required for end-to-end scientific simulations. \hl{Catching syntactic and executability errors is a necessary foundation, but closing the accuracy gap will require additional mechanisms; iteration within an agentic LLM harness offers one workable path.}

\hl{A consistent observation across the dataset is that errors} rarely happen in isolation. \hl{Scripts that contained} errors in pair\_style definitions, for example, \hl{also typically contained} additional inaccuracies in \hl{argument specifications, region definitions, or unit handling}. In more complex prompts, models also struggle with geometric reasoning, including consistent definition of regions, groups, and spatial relationships. These patterns suggest that the observed limitations arise from a combination of the complexity of the LAMMPS DSL, the limited availability of LAMMPS-specific training data, and the difficulty models face in reasoning over symbolic commands and physical constraints within a coherent workflow.

Aggregate success rates alone do not fully explain why LLMs fail on complex scientific workflows. \hl{The case studies below describe four representative failure mechanisms identified in our analysis}, progressing from errors in command selection to breakdowns in numerical grounding, semantic validity, and multi-constraint physical reasoning.

The most frequently observed failure mode was incorrect pair\_style \hl{specification}, affecting nearly a third of the \hl{150 scripts generated by older models and persisting in a smaller fraction of scripts produced by newer models}. This pattern reflects three compounding factors. First, embedded atom model (EAM) potentials are implemented in LAMMPS through multiple, closely related parameterizations, including `eam', `eam/alloy', `eam/fs', which differ subtly in usage despite sharing similar underlying mathematical formulations. Second, LAMMPS additionally supports EAM potential definitions through the OpenKIM \cite{Tadmor2011} database, introducing \hl{further} variability in how \hl{a potential} can be specified. Finally, the natural-language \hl{prompts} did not explicitly constrain which EAM variant to use, \hl{since selecting an} appropriate potential (e.g., `eam/alloy' for metallic systems) is typically treated as implicit domain knowledge rather than a user-specified instruction. 

\hl{A second} common failure \hl{was} the use of placeholder or default values in place of task-specific parameters. In several generated scripts, models reverted to generic values when required to infer material-dependent or unit-sensitive quantities from the prompt. One representative example is the specification of lattice parameters, which requires knowledge of the element(s) being simulated. When prompted to construct a crystal, models were expected to supply an appropriate lattice constant; however, in many cases it was set to 1 \AA, a generic value rather than a physically meaningful choice. \hl{Other parameters showed} similar behavior, including timestep and barostat or thermostat damping constants, which were occasionally set to default values or deviated from reasonable physical scales by orders of magnitude. As an example, in Prompt 3, projectile velocities were frequently specified as 2000 without an appropriate unit conversion, suggesting that some models implicitly assumed SI units (m/s) rather than the \AA/ps units required under `units metal'. This failure mode highlights a broader limitation: when required to translate physical quantities across units, scales, and material-specific properties, models often fail to perform the necessary contextual reasoning.

A third recurring error was the generation of command/argument pairings that appear syntactically plausible but are invalid in the LAMMPS DSL. In a subset of Prompt 3 scripts, models hallucinated a non-existent `velocity' style in which they used `velocity groupID add' to assign the impact velocity of the projectile, rather than the correct syntax `velocity groupID set vx vy vz sum yes'. This command variant is not supported by LAMMPS and does not appear in the official documentation\hl{, though we did find a} single reference to \hl{it} in an informal discussion forum \cite{ali2022velocity}, suggesting \hl{the form} reflects a natural way users might attempt to express \hl{an} incremental velocity assignment\hl{, even though the syntax is not valid. This example points to two related but distinct issues. From the user side, the LAMMPS DSL does not always provide direct syntax for the operations users want. From the model side, the hallucination reveals that models may extrapolate command semantics from perceived usage patterns rather than formal grammar, producing commands that look reasonable but fail validation. Hallucinations of this kind are difficult to catch by inspection and motivate the kind of structured grammar-grounded validation our parser is designed to provide}.

The most severe breakdowns were observed in Prompt 3, \hl{where} coordinating multiple interdependent assumptions \hl{is required to} construct a physically consistent simulation. \hl{Boundary conditions provide a clear example. The} prompt requests free boundaries \hl{along the shock direction, but the LAMMPS DSL implements these boundaries through two distinct options:} fixed or shrink-wrapped non-periodic boundaries. Although both choices are syntactically and formally valid, a fixed boundary would require extending the simulation box to accommodate high-velocity atoms, whereas a shrink-wrapped boundary would not. Selecting the appropriate option therefore requires anticipating downstream physical consequences rather than just applying a local syntax rule\hl{, exactly the kind of reasoning that becomes harder as the number of interdependent choices grows.}

The same multi-step reasoning is required for the problem of command sequencing, geometric setup, and unit translation. Distances in the prompt are expressed using a mixture of lattice units and nanometers, while the `units metal' command interprets all lengths in angstroms and requires the lattice constant to be inferred from the simulated material. These quantities must then be consistently translated into geometries and region definitions with correct command ordering. \hl{As an example of this type of failure mode, GPT-5.4 particularly struggled with command ordering, routinely placing run\_style in the script before the simulation box was defined, a critical error regardless of whether the underlying numerical values are correct.} Across models, \hl{the difficulty was not in handling any one of these elements in isolation, but in maintaining consistency across all of them at once}. While isolated components of the setup were often handled correctly, failures frequently emerged from inconsistencies across these coupled steps.

\hl{To test how these limitations manifest in a deployment setting, we analyzed the results from the agentic skill workflow described in Methods, with a domain expert reviewing each resulting script. The skill caught the localized failures it was designed to catch: across the six runs, the parser and reduced-step execution stages evaluated for the kinds of structural inconsistencies discussed in the case studies above. The Sonnet 4.6 Prompt 3 script, however, passed every automated stage of the skill but failed the human verification step on physical grounds, a result consistent with the argument that coupled failures cannot be resolved through structural checks alone, and that human verification and validation \mbox{\cite{thacker2004concepts}} remains a necessary step. Once a working script is generated, LLMs can accelerate this verification step by writing analysis routines for energy and momenta conservation or related sanity checks, but the judgment about which checks are appropriate for a given simulation still rests with the domain expert.}

\hl{Two behavioral patterns from the agentic runs are worth noting. Neither model used its web-search capability to consult LAMMPS documentation, relying instead on internal knowledge supplemented by the skill's reference guide, which encodes remediation patterns from our pipeline experiments. This suggests that documentation-grounded patterns may be more useful when included as invocable skills, rather than when accessible as a more general separate capability. Both models also performed minimal iteration within the skill, meaning the tool functioned more as a pre-execution check rather than an active repair loop for these scenarios. As models continue to improve, the value of the validation tooling will shift from catching obvious errors to identifying the subtler ones that models would otherwise miss.}

\hl{Failure modes in this study divide into two structurally different categories with different prospects for tooling-based mitigation. Our prompts} already expose substantial limitations, \hl{but} they remain simpler than many simulations routinely performed by MD practitioners, suggesting these challenges will be amplified in realistic research settings. \hl{Many failures at lower complexity} are localized and structurally identifiable (mismatched command usage, incomplete argument specification\hl{, hallucinated syntax) and domain-aware validation can detect these} errors, enforce structural constraints, and support iterative refinement. Failures in more complex workflows\hl{, in contrast,} arise from inconsistencies spanning multiple interdependent steps \hl{and} cannot be resolved through syntax or structural checks alone. \hl{This is a pattern visible in both the one-shot study and the agentic runs, where the skill caught structural failures cleanly but coupled failures still required human verification. The model performance data we report reflects this division: parsing and execution rates improve substantially across model generations, but accuracy in the most complex prompt remains low even in the new models. State-of-the-art LLMs are therefore best deployed as} assistive components within workflows that combine generation with validation, feedback, and expert oversight.\hl{ As models continue to improve, structured evaluation tools like the one introduced here will be essential for tracking that progress.}

\section*{Future Work}
\label{future_work_section}

Translating natural language physics descriptions into reliable MD simulations requires \hl{more than} runnable code\hl{; it requires scripts that }capture the intended physical and scientific constraints \cite{verduzco2023gpt}. In scientific DSLs such as LAMMPS, correctness is not determined solely by syntax but also by the consistent handling of units, geometries, boundary conditions, and material-specific parameters, as well as by the semantically correct ordering of commands. \hl{Our results show that current LLMs produce syntactically valid scripts with growing reliability but still fall short of this fuller notion of correctness, particularly on coupled multi-step workflows. The agentic skill demonstrated here begins to close that gap by delivering structured validation feedback inside the generation loop, and the small-scale results suggest this is a productive direction. Several open challenges remain, however, before such workflows can be considered routinely reliable: assessing scientific correctness automatically, characterizing agentic behavior at scale, growing the validation infrastructure with the community, and integrating generation with FAIR simulation-data ecosystems.}

\hl{The most pressing limitation of the current pipeline is the absence of automated mechanisms for assessing scientific correctness. Our accuracy stage relies on prompt-specific parameter checklists scored manually by domain experts, which is reliable but does not scale and does not generalize easily across simulation types. A better approach would be to monitor simulation behavior directly: energy and momentum conservation, equilibration of temperatures and pressures to their targets, density evolution during melting, and other similar property-level checks of the kind researchers already perform after a run. These checks can typically run cheaply on reduced-step or downsampled simulations and require no per-prompt configuration. They also offer a path to address the cost gap: by catching physical inconsistencies early through inexpensive signals rather than full production runs, the iteration loop becomes affordable enough for an agent to use repeatedly during script repair, and will, ideally reduce the prevalence of poor-quality production runs, which represent a costly, often ignored facet of MD research.}

\hl{The agentic skill experiments in this work were limited to six runs across two models, large enough to demonstrate the approach but too small to characterize its behavior in detail. A more systematic study would expand sample sizes, sweep across additional models, and identify the conditions under which the skill drives iterative repair rather than functioning only as a pre-execution check. Comparing agentic and one-shot generation against these same prompts would also quantify the cost-performance tradeoff, as agentic runs currently result in roughly an order of magnitude larger token usage. Reasoning shortcuts, including curated libraries of LAMMPS idioms callable as part of the skill, documentation-grounded retrieval to reduce hallucinated command usage, and structured remediation patterns derived from common failure modes, would substantially reduce both token usage and repair iterations. The goal is to make iterative validation accessible enough that it becomes a default rather than a luxury.}

\hl{The evaluation pipeline and parser introduced in this work are open-source and designed for community extension. Expanding its command coverage toward the full LAMMPS grammar, supporting differences across versions, and lowering the barrier to submit grammar contributions are concrete steps towards broader applicability. This matters most for system types underrepresented in the current study: molecular, reactive, and multi-component simulations, which carry more implicit physical and structural assumptions. A community-maintained grammar avoids the maintenance bottleneck of a single research group while keeping the validation logic transparent and inspectable, in contrast to validation embedded inside an opaque model. Beyond validation, a complete LAMMPS grammar would enable structure-aware approaches further upstream:} recent work has explored incorporating structural representations such as ASTs directly into model pre-training to improve syntactic reliability in code generation \mbox{\cite{Gong2024}}. Combining structure-aware pretraining with downstream validation and execution-aware feedback \hl{is} a promising direction for scientific DSL pipelines.

\begin{figure}[h!]
  \centering
  \includegraphics[width=.98\textwidth]{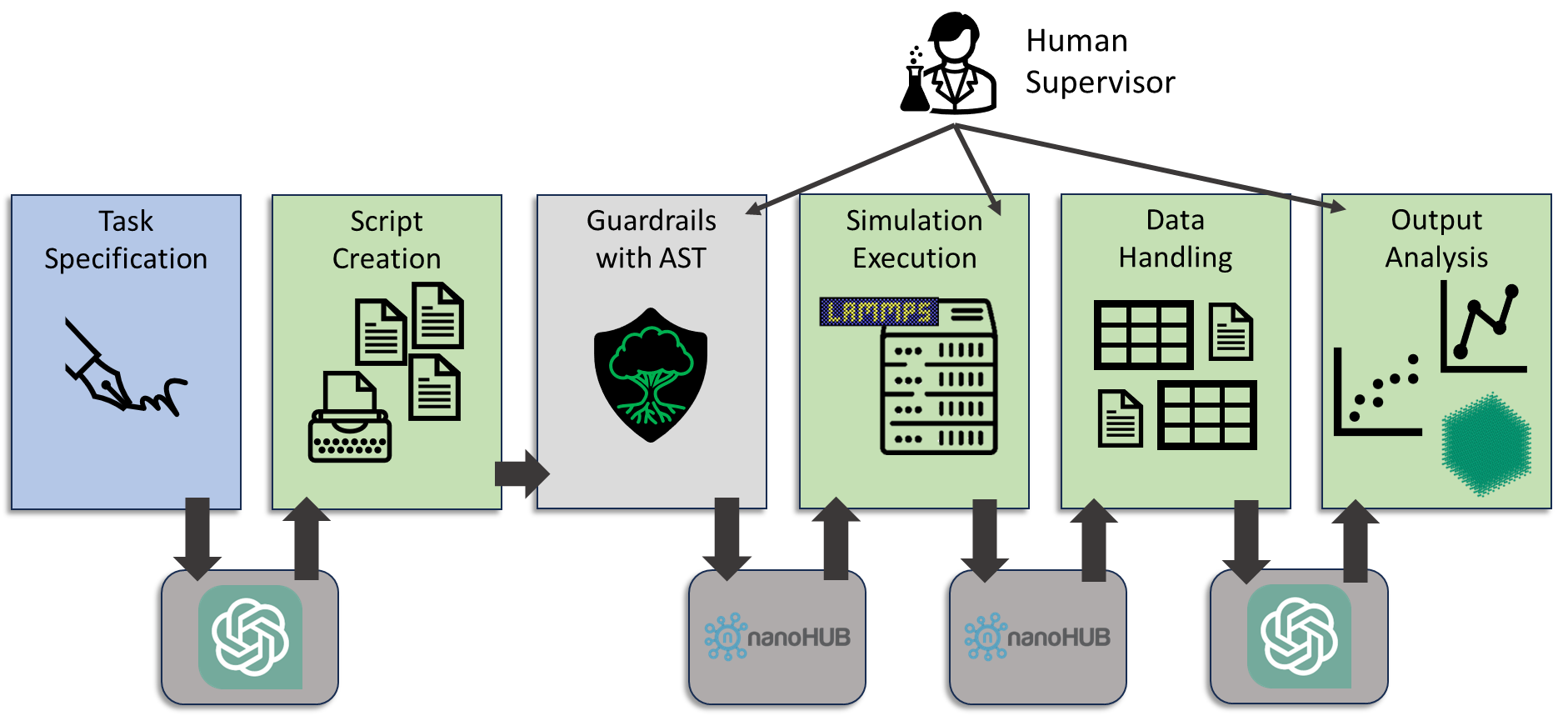}
  \caption{Towards an autonomous workflow. A prototype pipeline implemented on nanoHUB integrates LLM-driven script generation, AST-based validation, simulation execution, and data handling, with a domain expert overseeing scientific validation and analysis stages.}
  \label{fig:Workflow}
\end{figure}

\hl{As AI-generated scripts move from experimental curiosity to routine use, the metadata around their generation becomes an important research artifact. A single simulation script now carries a generation history including the model that generated the script, the prompts used, the intermediate repair steps, and the validation outcomes. This metadata is valuable for reproducibility, debugging, and analysis of model behavior. Capturing this generation-to-result chain as structured provenance, aligned with FAIR data principles \mbox{\cite{wilkinson2016fair}} would let researchers query not just simulations but the processes that produced them. As a step in this direction}, we have developed a prototype workflow that integrates script generation, validation, and execution into a unified research pipeline, as illustrated in Figure \ref{fig:Workflow}.

\hl{The workflow is implemented as} a Sim2l \cite{Hunt2022} published on nanoHUB \cite{strachan2010cyber} that leverages the platform's computational infrastructure to execute LAMMPS simulations and archive resulting data. The tool integrates two LLM-powered code-generation components: LAMMPS input script synthesis and Python-based post-processing and visualization. The static parser is integrated as a pre-execution validation layer, ensuring that generated scripts satisfy syntax and command-level correctness before simulation submission. \hl{Each cycle of the pipeline is recorded alongside the simulation output, producing the kind of structured provenance described above. Expert oversight is still required for scientific validation, but the surrounding infrastructure for capturing, archiving, and querying AI-generated simulation workflows is in place.}

\section*{Conclusions}
\label{conclusions_section}

In this work, we introduce a procedure to evaluate LAMMPS input scripts generated by language models from natural language descriptions, applied to eight state-of-the-art LLMs across three prompts of increasing complexity. While the models tested frequently produce syntactically plausible and partially correct scripts, performance and reliability degrade sharply as simulation workflows increase in structural and physical complexity. Common failure modes include incorrect pair style selection, inconsistent unit handling, malformed command arguments, and breakdowns in multi-step physical reasoning. \hl{Within the scope of the workflows tested, our pipeline reveals that current LLMs in their base configurations are not yet reliable as fully autonomous scientific simulation designers.}
Newer models \hl{substantially} improve \hl{structural reliability, but scientific accuracy on the most complex prompt remains low, and} as outputs \hl{become more plausible}, undetected errors \hl{become} increasingly difficult to catch. \hl{Systematic evaluation tools therefore matter more, not less, as models improve.}

\hl{Packaging the evaluation pipeline as an agentic skill begins to close the accuracy gap: in our small-scale demonstration, two recent models produced five fully accurate scripts out of six across all three prompts, including the most complex one. This suggests that structured validation feedback delivered inside the generation loop is a productive direction, though human verification remains necessary for the physical correctness checks that structural tools cannot capture.} By explicitly separating structural correctness from physical validity, \hl{our work} establishes a foundation for AI-assisted simulation pipelines \hl{that preserve} reliability and reproducibility  \hl{without relying on the model alone. The parser and skill are open-source and designed for community extension, providing a path for the validation infrastructure to grow alongside both the LAMMPS ecosystem and the language models that interact with it. The automation of LAMMPS simulations is increasingly a question of building the right guardrails and verification methodology, not of raw model capability.} 

\section*{Code and Data Availability}
\label{availability_section}

An interactive tool for the \href{https://nanohub.org/tools/llm4lammps}{LAMMPS Parser}\cite{llm4lammps} is available for online simulations on nanoHUB.org\cite{strachan2010cyber}.\hl{All code and scripts used for the evaluation pipeline are publicly available at} \url{https://github.com/ethanholbrook/LAMMPS-AST}, \hl{including the grammar file, normalization routines, and evaluation checklists. The parser is also available as a Python package via \texttt{pip install lammps-ast}.}

\section*{Acknowledgements}

This effort was supported by the US National Science Foundation FAIROS program, award 2226418.

\section*{Ethics declarations}

\subsection*{Conflict of interest}

On behalf of all authors, the corresponding author states that there is no conflict of interest.

\subsection*{Declaration of generative AI and AI-assisted technologies in the manuscript preparation process}

 During the preparation of this work the authors used chatGPT, Claude, Codex, and Claude Code in order to develop the coding pipeline, refine manuscript wording, and generate LAMMPS input scripts. After using this tool/service, the authors reviewed and edited the content as needed and take full responsibility for the content of the published article.

\subsection*{Author Information}

\textbf{Corresponding Author}

\textbf{Alejandro Strachan} - School of Materials Engineering and Birck Nanotechnology Center, Purdue University, West Lafayette, Indiana 47907; Email: strachan@purdue.edu \\

\bibliographystyle{unsrt}
\bibliography{references.bib}

@techreport{thacker2004concepts,
  title={Concepts of model verification and validation},
  author={Thacker, Ben H and Doebling, Scott W and Hemez, Francois M and Anderson, Mark C and Pepin, Jason E and Rodriguez, Edward A},
  year={2004},
  publisher={Los Alamos National Lab., Los Alamos, NM (US)},
  institution = {Los Alamos National Lab}
}

@article{lammps_tutorials_2025,
  author={Gravelle, Simon and Alvares, Cecilia M. S. and Gissinger, Jacob R. and Kohlmeyer, Axel},
  title={A Set of Tutorials for the {LAMMPS} Simulation Package [Article v1.0]},
  journal={Living Journal of Computational Molecular Science},
  pages={3027},
  volume={6},
  number={1},
  year={2025},
  month={Sep.},
  url={https://livecomsjournal.org/index.php/livecoms/article/view/v6i1e3037},
  DOI={10.33011/livecoms.6.1.3037}
}

@article{thompson2022lammps,
  title={LAMMPS-a flexible simulation tool for particle-based materials modeling at the atomic, meso, and continuum scales},
  author={Thompson, Aidan P and Aktulga, H Metin and Berger, Richard and Bolintineanu, Dan S and Brown, W Michael and Crozier, Paul S and in't Veld, Pieter J and Kohlmeyer, Axel and Moore, Stan G and Nguyen, Trung Dac and others},
  journal={Computer Physics Communications},
  volume={271},
  pages={108171},
  year={2022},
  publisher={Elsevier}
}

@techreport{openai2025gpt5systemcard,
  author       = {{OpenAI}},
  title        = {{GPT-5 System Card [Large language model]}},
  year         = {2025},
  month        = {August},
  institution  = {{OpenAI}},
  url          = {https://cdn.openai.com/gpt-5-system-card.pdf},
  note         = {Accessed: 2025-10-07}
}

@techreport{anthropic2025claudesystemcard,
  author       = {Anthropic},
  title        = {{System Card: Claude Opus 4 \& Claude Sonnet 4 [Large language model]}},
  year         = {2025},
  month        = {May},
  institution  = {Anthropic},
  url          = {https://www-cdn.anthropic.com/4263b940cabb546aa0e3283f35b686f4f3b2ff47.pdf},
  note         = {Accessed: 2025-10-07}
}

@techreport{deepmind2025gemini_2_5_report,
  author       = {{DeepMind / Google}},
  title        = {{Gemini 2.5: Pushing the Frontier with Advanced Reasoning [Large language model]}},
  year         = {2025},
  institution  = {{DeepMind / Google}},
  url          = {https://storage.googleapis.com/deepmind-media/gemini/gemini_v2_5_report.pdf},
  note         = {Accessed: 2025-10-07}
}

@inproceedings{xia2023automated,
  title={Automated program repair in the era of large pre-trained language models},
  author={Xia, Chunqiu Steven and Wei, Yuxiang and Zhang, Lingming},
  booktitle={2023 IEEE/ACM 45th International Conference on Software Engineering (ICSE)},
  pages={1482--1494},
  year={2023},
  organization={IEEE}
}

@article{boiko2023autonomous,
  title={Autonomous chemical research with large language models},
  author={Boiko, Daniil A and MacKnight, Robert and Kline, Ben and Gomes, Gabe},
  journal={Nature},
  volume={624},
  number={7992},
  pages={570--578},
  year={2023},
  publisher={Nature Publishing Group UK London}
}

@article{m2024augmenting,
  title={Augmenting large language models with chemistry tools},
  author={M. Bran, Andres and Cox, Sam and Schilter, Oliver and Baldassari, Carlo and White, Andrew D and Schwaller, Philippe},
  journal={Nature Machine Intelligence},
  volume={6},
  number={5},
  pages={525--535},
  year={2024},
  publisher={Nature Publishing Group UK London}
}

@article{ghafarollahi2025automating,
  title={Automating alloy design and discovery with physics-aware multimodal multiagent AI},
  author={Ghafarollahi, Alireza and Buehler, Markus J},
  journal={Proceedings of the National Academy of Sciences},
  volume={122},
  number={4},
  pages={e2414074122},
  year={2025},
  publisher={National Academy of Sciences}
}

@article{wang2025dreams,
  title={DREAMS: Density Functional Theory Based Research Engine for Agentic Materials Simulation},
  author={Wang, Ziqi and Huang, Hongshuo and Zhao, Hancheng and Xu, Changwen and Zhu, Shang and Janssen, Jan and Viswanathan, Venkatasubramanian},
  journal={arXiv preprint arXiv:2507.14267},
  year={2025}
}

@article{joel2024survey,
  title={A survey on llm-based code generation for low-resource and domain-specific programming languages},
  author={Joel, Sathvik and Wu, Jie JW and Fard, Fatemeh H},
  journal={arXiv preprint arXiv:2410.03981},
  year={2024}
}

@article{jacobs2025developing,
  title={Developing large language models for quantum chemistry simulation input generation},
  author={Jacobs, Pieter Floris and Pollice, Robert},
  journal={Digital Discovery},
  volume={4},
  number={3},
  pages={762--775},
  year={2025},
  publisher={Royal Society of Chemistry}
}

@article{mudur2025feabench,
  title={FEABench: Evaluating Language Models on Multiphysics Reasoning Ability},
  author={Mudur, Nayantara and Cui, Hao and Venugopalan, Subhashini and Raccuglia, Paul and Brenner, Michael P and Norgaard, Peter},
  journal={arXiv preprint arXiv:2504.06260},
  year={2025}
}

@article{mendible2025dynamate,
  title={DynaMate: leveraging AI-agents for customized research workflows},
  author={Mendible-Barreto, Orlando A and D{\'\i}az-Maldonado, Misael and Esteva, Fernando J Carmona and Torres, J Emmanuel and C{\'o}rdova-Figueroa, Ubaldo M and Col{\'o}n, Yamil J},
  journal={Molecular Systems Design \& Engineering},
  year={2025},
  publisher={Royal Society of Chemistry}
}

@article{shi2025fine,
  title={A fine-tuned large language model based molecular dynamics agent for code generation to obtain material thermodynamic parameters},
  author={Shi, Zhuofan and Xin, Chunxiao and Huo, Tong and Jiang, Yuntao and Wu, Bowen and Chen, Xingyue and Qin, Wei and Ma, Xinjian and Huang, Gang and Wang, Zhenyu and others},
  journal={Scientific Reports},
  volume={15},
  number={1},
  pages={10295},
  year={2025},
  publisher={Nature Publishing Group UK London}
}

@article{wei2022chain,
  title={Chain-of-thought prompting elicits reasoning in large language models},
  author={Wei, Jason and Wang, Xuezhi and Schuurmans, Dale and Bosma, Maarten and Xia, Fei and Chi, Ed and Le, Quoc V and Zhou, Denny and others},
  journal={Advances in neural information processing systems},
  volume={35},
  pages={24824--24837},
  year={2022}
}

@article{kojima2022large,
  title={Large language models are zero-shot reasoners},
  author={Kojima, Takeshi and Gu, Shixiang Shane and Reid, Machel and Matsuo, Yutaka and Iwasawa, Yusuke},
  journal={Advances in neural information processing systems},
  volume={35},
  pages={22199--22213},
  year={2022}
}

@article{verduzco2023gpt,
  title={GPT-4 as an interface between researchers and computational software: improving usability and reproducibility},
  author={Verduzco, Juan C and Holbrook, Ethan and Strachan, Alejandro},
  journal={arXiv preprint arXiv:2310.11458},
  year={2023}
}

@misc{Lark2025,
  author       = {Shinan, Erez},
  title        = {Lark: A parsing toolkit for Python},
  howpublished = {\url{https://github.com/lark-parser/lark}},
  note         = {Version 1.3.1},
  year         = {2025}
}

@article{Mishin1999,
author = {Mishin, Y. and Farkas, D. and Mehl, M. J. and Papaconstantopoulos, D. A.},
doi = {10.1103/PhysRevB.59.3393},
issn = {1550235X},
journal = {Physical Review B - Condensed Matter and Materials Physics},
number = {5},
pages = {3393--3407},
title = {{Interatomic potentials for monoatomic metals from experimental data and ab initio calculations}},
volume = {59},
year = {1999}
}

@article{Tadmor2011,
author = {Tadmor, E. B. and Elliott, R. S. and Sethna, J. P. and Miller, R. E. and Becker, C. A.},
doi = {10.1007/s11837-011-0102-6},
issn = {10474838},
journal = {Jom},
number = {7},
pages = {17},
title = {{The potential of atomistic simulations and the knowledgebase of interatomic models}},
volume = {63},
year = {2011}
}

@article{Hunt2022,
archivePrefix = {arXiv},
arxivId = {2110.06886},
author = {Hunt, Martin and Clark, Steven and Mejia, Daniel and Desai, Saaketh and Strachan, Alejandro},
doi = {10.1371/journal.pone.0264492},
eprint = {2110.06886},
isbn = {1111111111},
issn = {19326203},
journal = {PLoS ONE},
number = {3 March},
pages = {1--14},
pmid = {35271613},
title = {{Sim2Ls: FAIR simulation workflows and data}},
url = {http://dx.doi.org/10.1371/journal.pone.0264492},
volume = {17},
year = {2022}
}

@article{strachan2010cyber,
title={Cyber-enabled simulations in nanoscale science and engineering},
author={Strachan, Alejandro and Klimeck, Gerhard and Lundstrom, Mark},
journal={Computing in Science \& Engineering},
volume={12},
number={2},
pages={12--17},
year={2010},
publisher={IEEE}
}

@article{Gu2025,
   author = {Xiaodong Gu and Meng Chen and Yalan Lin and Yuhan Hu and Hongyu Zhang and Chengcheng Wan and Zhao Wei and Yong Xu and Juhong Wang},
   doi = {10.1145/3697012},
   issn = {15577392},
   issue = {3},
   journal = {ACM Transactions on Software Engineering and Methodology},
   month = {2},
   publisher = {Association for Computing Machinery},
   title = {On the Effectiveness of Large Language Models in Domain-Specific Code Generation},
   volume = {34},
   year = {2025}
}

@article{wang2023grammar,
  title={Grammar prompting for domain-specific language generation with large language models},
  author={Wang, Bailin and Wang, Zi and Wang, Xuezhi and Cao, Yuan and A Saurous, Rif and Kim, Yoon},
  journal={Advances in Neural Information Processing Systems},
  volume={36},
  pages={65030--65055},
  year={2023}
}

@article{Gong2024,
archivePrefix = {arXiv},
arxivId = {2401.03003},
author = {Gong, Linyuan and Elhoushi, Mostafa and Cheung, Alvin},
eprint = {2401.03003},
issn = {26403498},
journal = {Proceedings of Machine Learning Research},
pages = {15839--15853},
title = {{AST-T5: Structure-Aware Pretraining for Code Generation and Understanding}},
volume = {235},
year = {2024}
}

@article{Phillips2020,
author = {Phillips, James C. and Hardy, David J. and Maia, Julio D.C. and Stone, John E. and Ribeiro, Jo{\~{a}}o V. and Bernardi, Rafael C. and Buch, Ronak and Fiorin, Giacomo and H{\'{e}}nin, J{\'{e}}r{\^{o}}me and Jiang, Wei and McGreevy, Ryan and Melo, Marcelo C.R. and Radak, Brian K. and Skeel, Robert D. and Singharoy, Abhishek and Wang, Yi and Roux, Beno{\^{i}}t and Aksimentiev, Aleksei and Luthey-Schulten, Zaida and Kal{\'{e}}, Laxmikant V. and Schulten, Klaus and Chipot, Christophe and Tajkhorshid, Emad},
doi = {10.1063/5.0014475},
issn = {10897690},
journal = {Journal of Chemical Physics},
number = {4},
pmid = {32752662},
publisher = {AIP Publishing, LLC},
title = {{Scalable molecular dynamics on CPU and GPU architectures with NAMD}},
url = {https://doi.org/10.1063/5.0014475},
volume = {153},
year = {2020}
}

@article{Kresse1996,
author = {Kresse, G. and Furthm{\"{u}}ller, J.},
doi = {10.1103/PhysRevB.54.11169},
issn = {0163-1829},
journal = {Physical Review B},
month = {oct},
number = {16},
pages = {11169--11186},
pmid = {13656531},
title = {{Efficient iterative schemes for ab initio total-energy calculations using a plane-wave basis set}},
url = {https://link.aps.org/doi/10.1103/PhysRevB.54.11169},
volume = {54},
year = {1996}
}

@article{Giannozzi_2009,
doi = {10.1088/0953-8984/21/39/395502},
url = {https://doi.org/10.1088/0953-8984/21/39/395502},
year = {2009},
month = {sep},
publisher = {},
volume = {21},
number = {39},
pages = {395502},
author = {Giannozzi, Paolo and Baroni, Stefano and Bonini, Nicola and Calandra, Matteo and Car, Roberto and Cavazzoni, Carlo and Ceresoli, Davide and Chiarotti, Guido L and Cococcioni, Matteo and Dabo, Ismaila and Dal Corso, Andrea and de Gironcoli, Stefano and Fabris, Stefano and Fratesi, Guido and Gebauer, Ralph and Gerstmann, Uwe and Gougoussis, Christos and Kokalj, Anton and Lazzeri, Michele and Martin-Samos, Layla and Marzari, Nicola and Mauri, Francesco and Mazzarello, Riccardo and Paolini, Stefano and Pasquarello, Alfredo and Paulatto, Lorenzo and Sbraccia, Carlo and Scandolo, Sandro and Sclauzero, Gabriele and Seitsonen, Ari P and Smogunov, Alexander and Umari, Paolo and Wentzcovitch, Renata M},
title = {QUANTUM ESPRESSO: a modular and open-source software project for quantum
simulations of materials},
journal = {Journal of Physics: Condensed Matter}
}

@article{harbour2025moose,
   title = {4.0 {MOOSE}: Enabling massively parallel Multiphysics simulation},
 journal = {SoftwareX},
  volume = {31},
   pages = {102264},
    year = {2025},
    issn = {2352-7110},
     doi = {https://doi.org/10.1016/j.softx.2025.102264},
     url = {https://www.sciencedirect.com/science/article/pii/S2352711025002316},
  author = {Logan Harbour and Guillaume Giudicelli and Alexander D. Lindsay and Peter German and
            Joshua Hansel and Casey Icenhour and Mengnan Li and Jason M. Miller and Roy H. Stogner and
            Patrick Behne and Daniel Yankura and Zachary M. Prince and Corey DeChant and Daniel Schwen and
            Benjamin W. Spencer and Mauricio Tano and Namjae Choi and Yaqi Wang and Max Nezdyur and
            Yinbin Miao and Tianchen Hu and Shikhar Kumar and Christopher Matthews and Brandon Langley and
            Nuno Nobre and Alexander Blair and Chris MacMackin and Henrique Bergallo Rocha and
            Edward Palmer and Jesse Carter and J{\"o}rg Meier and Andrew E. Slaughter and David Andr{\v{s}} and
            Robert W. Carlsen and Fande Kong and Derek R. Gaston and Cody J. Permann}
}

@article{wilkinson2016fair,
  title={The FAIR Guiding Principles for scientific data management and stewardship},
  author={Wilkinson, Mark D and Dumontier, Michel and Aalbersberg, IJsbrand Jan and Appleton, Gabrielle and Axton, Myles and Baak, Arie and Blomberg, Niklas and Boiten, Jan-Willem and da Silva Santos, Luiz Bonino and Bourne, Philip E and others},
  journal={Scientific data},
  volume={3},
  number={1},
  pages={1--9},
  year={2016},
  publisher={Nature Publishing Group}
}

@misc{ali2022velocity,
  author       = {Ali, Muhammad Saad and Gravelle, Simon and Kohlmeyer, Axel},
  title        = {Velocity Equilibration Problem},
  year         = {2022},
  month        = nov,
  howpublished = {Materials Science Community Discourse (matsci.org), LAMMPS Beginners},
  url          = {https://matsci.org/t/velocity-equilibration-problem/45571},
  note         = {Online forum thread discussing LAMMPS velocity command usage; accessed March 18, 2026}
}

@misc {llm4lammps,
	title = {Large Language model demonstration for LAMMPS},
	month = {Jan},
	url = {https://nanohub.org/resources/llm4lammps},
	year = {2024},
	doi = {doi:10.21981/TXS6-5C79},
	author = {Holbrook , Ethan and Gastelum , Juan Carlos Verduzco and Mishra , Saswat and Nykiel , Kat and Zummo , William and Strachan , Alejandro}
}

@software{claudecode2025,
  author = {Anthropic},
  title = {Claude Code},
  year = {2025},
  url = {https://claude.ai/code},
  note = {Accessed: 2025}
}

@misc{zhang2025agentskills,
  author       = {Zhang, Barry and Lazuka, Keith and Murag, Mahesh},
  title        = {Equipping Agents for the Real World with Agent Skills},
  year         = {2025},
  month        = {October},
  day          = {16},
  howpublished = {\url{https://www.anthropic.com/engineering/equipping-agents-for-the-real-world-with-agent-skills}},
  note         = {Engineering at Anthropic. Accessed: May 13, 2026},
  organization = {Anthropic}
}

@article{Shi2026,
archivePrefix = {arXiv},
arxivId = {2601.02075},
author = {Shi, Zhuofan and A, Hubao and Shao, Yufei and Huang, Dongliang and An, Hongxu and Xin, Chunxiao and Shen, Haiyang and Wang, Zhenyu and Na, Yunshan and Huang, Gang and Jing, Xiang},
eprint = {2601.02075},
month = {feb},
title = {{MDAgent2: Large Language Model for Code Generation and Knowledge Q\&A in Molecular Dynamics}},
url = {http://arxiv.org/abs/2601.02075},
year = {2026},
journal = {arXiv}
}

\renewcommand{\thefigure}{S\arabic{figure}}
\renewcommand{\thetable}{S\arabic{table}}

\setcounter{figure}{0}
\setcounter{table}{0}
\newpage
\section*{Supplemental Information: Evaluating LLM-generated code for domain-specific languages: molecular dynamics with LAMMPS}

\subsection*{Inference Settings and Estimated Costs}
All models were accessed via their respective public APIs using default parameters. No fine-tuning, temperature adjustment, or other parameter modifications were applied unless noted. A total of 240 generations were produced across eight models and three prompts, with ten independent generations per prompt/model combination. Table \ref{tab:costs} summarizes the model identifiers, token usage, and estimated costs associated with each model.

\begin{table}[h]
\centering
\caption{Inference costs and token usage for scripts across all models, 
as reported by provider billing dashboards.}
\label{tab:costs}
\sisetup{group-separator={,}, group-minimum-digits=4}
\begin{tabular}{l
                S[table-format=5.0]
                S[table-format=6.0]
                rrr}
\toprule
Model & {Input tokens} & {Output tokens} & Input (USD) & Output (USD) & Total (USD) \\
\midrule
GPT-4o     & 12847 & 4381 & \$0.03 & \$0.04 & \$0.07 \\
GPT-4.1    & 16760 & 2585 & \$0.07 & \$0.07 & \$0.14 \\
o3         & 14212 & 50198 & \$0.06 & \$1.27 & \$1.33 \\
GPT-5      & 20810 & 210486 & \$0.03 & \$2.10 & \$2.13 \\
Claude Opus 4   & 24173 & 7379  & \$0.36 & \$0.56 & \$0.92 \\
GPT-5.4 (High)  & 38533 & 369079 & \$0.10 & \$5.54 & \$5.64 \\
Claude Opus 4.7 & 23530 & 8442 & \$0.12 & \$0.33 & \$0.45 \\
GPT-5.5         & 14920 & 113591 & \$0.07 & \$3.41 & \$3.48 \\
\bottomrule
\end{tabular}
\end{table}

To assess the agentic variant of this pipeline, we executed each of the three prompts once using the agentic skill version with two models (Claude Opus 4.7, Claude Sonnet 4.6), rather than the ten independent generations across eight models used in the non-agentic setting. Because agentic execution requires substantially more tokens per run due to tool calls, intermediate reasoning, and multi-turn interactions, we limited this evaluation to a single generation per prompt. Table~\ref{tab:agentic-tokens} reports the per-prompt token usage, broken down into fresh input, output, and prompt-cache reads and writes; cache reads dominate total volume, as the system prompt and accumulated context are re-read on each turn of the agentic loop.

\begin{table}[h]
\centering
\caption{Token usage for the agentic skill variant across three prompts, 
run once per prompt. Cache reads dominate total input volume, as expected 
for multi-turn agentic workloads with a stable system prompt.}
\label{tab:agentic-tokens}
\sisetup{group-separator={,}, group-minimum-digits=4}
\begin{tabular}{ll
                S[table-format=2.0]
                S[table-format=3.0]
                S[table-format=6.0]
                S[table-format=7.0]
                S[table-format=6.0]}
\toprule
Model & Prompt & {Turns} & {Input} & {Output} & {Cache read} & {Cache write} \\
\midrule
\multirow{3}{*}{Opus 4.7} 
 & Prompt 1 & 44 & 66 & 17674 & 778422 & 50615 \\
 & Prompt 2 & 43 & 62 & 15344 & 741424 & 55807 \\
 & Prompt 3 & 44 & 67 & 38048 & 912738 & 75642  \\
\midrule
\multirow{3}{*}{Sonnet 4.6} 
 & Prompt 1 & 60 & 53 & 15889 & 841096  & 51622 \\
 & Prompt 2 & 52 & 53 & 16802 & 787790  & 52900 \\
 & Prompt 3 & 55 & 52 & 63948 & 1159867 & 70607 \\
\bottomrule
\end{tabular}
\end{table}

\newpage

\subsection*{Prompting}
The system prompt \hl{for the pipeline} is included below, \hl{formatted for human-readability from the raw string native to python API inputs:} 
\begin{promptquotebox}{System Prompt}

You are an expert in molecular dynamics simulations and LAMMPS scripting.
Your task is to generate complete, runnable LAMMPS input scripts based only
on the provided method description. You must follow strict formatting rules
and proceed step by step, reasoning through each section logically before
writing the final output. All reasoning should be internal---do not display
intermediate thoughts, summaries, or explanations. Output only the final script.

\medskip
\textbf{Input Format}

You will receive a user input labeled \textit{Method Description} containing
all experimental and simulation details.

\medskip
\textbf{Chain-of-Thought Workflow}

\begin{enumerate}[noitemsep]
    \item Parse and interpret the method description carefully. Extract only
    information explicitly stated; do not assume any values or simulation
    parameters beyond what is given.

    \item Plan the LAMMPS input script using the following clearly labeled sections:
    \begin{itemize}[noitemsep]
        \item Initialization
        \item System Construction
        \item Potential
        \item Miscellaneous (if needed)
        \item Production Run
    \end{itemize}

    \item Generate each section by:
    \begin{itemize}[noitemsep]
        \item Determining what LAMMPS commands are required based on the
        method description.
        \item Using default values only when the method explicitly mentions
        them or when required by LAMMPS syntax, but still writing them explicitly.
        \item Ensuring syntax correctness and functional completeness.
    \end{itemize}
\end{enumerate}

\textbf{Constraints}

\begin{itemize}[noitemsep]
    \item Assume a potential file named
    \texttt{../../../potentials/prompt1.potential} is available and ready to use.
    \item Select the pair style that specifically matches the format and type of
    the potential described. If a citation or filename indicates a specific
    format, infer the correct \texttt{pair\_style} accordingly
    (e.g., \texttt{eam/alloy}, \texttt{meam}, \texttt{tersoff}, \texttt{reaxff}).
    \item Do not include any comments in the script.
    \item Explicitly define all commands, even when using defaults.
    \item Output a single clean script, correctly structured and ready to run,
    with no explanations, metadata, or commentary.
\end{itemize}

\hl{
\textbf{Method Description:}

\textless Task Prompt \textgreater
}
\end{promptquotebox}
\newpage
The individual prompts are shown below: 
\begin{promptquotebox}{Prompt: Task 1}

We used molecular dynamics with LAMMPS to equilibrate an Al sample under isobaric, isothermal conditions (NPT ensemble) at 300 K and 1 atm for 500 ps. The initial conditions were obtained by replicating the fcc unit cell 5 times in each direction. We used Nose-Hoover thermostat and barostat with relaxation timescales of 0.1 and 1 ps, respectively. All MD simulations were performed using LAMMPS. Atomic interactions were obtained using an embedded atom model developed by Ercolessi and Adams [1] obtained from OpenKIM.org. [1] EAM alloy potential for Al developed by Ercolessi and Adams (1994) v002. OpenKIM. 2018. doi:10.25950/376e3e7e.
\end{promptquotebox}

\begin{promptquotebox}{Prompt: Task 2}

We characterized the melting of a bulk Ni sample using molecular dynamics with LAMMPS. The initial condition was obtained by replicating the Ni unit cell 10 times in each direction. Initial velocities were drawn from the Maxwell-Boltzmann distribution at 600 K. The system was heated from 300 K to 2500 K continuously, at a rate of 10 K per ps under isothermal and isobaric conditions at 1 atm. Interactions were described using an embedded atom model developed by Mishin et al. in 1999 [1] obtained from OpenKIM.org. [1] EAM potential (LAMMPS cubic hermite tabulation) for Ni developed by Mishin et al. (1999) v005. OpenKIM; 2018. doi:10.25950/a88dfc37.
\end{promptquotebox}

\begin{promptquotebox}{Prompt: Task 3}

We simulate spall failure on Nb single crystals using high-velocity impact simulations using molecular dynamics (MD) with the LAMMPS code. We simulate the impact of a projectile on a target with a relative velocity of 2 km/s. The projectile is obtained by replicating the Nb BCC unit cell 20 times along the [100], [010], and [001]; the target is longer along the shock direction and is obtained by replicating the BCC unit cell 20 times along [100] and [010] and 40 times along [001]. We apply periodic boundary conditions along the directions normal to the impact direction, [001], and free boundaries along [001]. A gap of 1.5 nm initially separates the target and projectile. The system is equilibrated at 300 K for 100 ps using isothermal, isochoric MD. An impact velocity of 2 km/s is added to the thermal velocities to all the atoms in the projectile along [001] in the direction of the target. Adiabatic MD is used to simulate the impact and subsequent expansion. All atomic interactions are described using an EAM potential developed by Fellinger et al. [1] and downloaded from OpenKIM [2]. [1] Fellinger MR, Park H, Wilkins JW. Force-matched embedded-atom method potential for niobium. Physical Review B. 2010Apr;81(14):144119. doi:10.1103/PhysRevB.81.144119 [2] https://doi.org/10.25950/befb2eea.
\end{promptquotebox}

\newpage
The prompt for the Agentic Skill is very minimal, but it is included below for reproducibility: 
\begin{promptquotebox}{Agent Prompt}

Generate and evaluate a LAMMPS script for the method description below, potentials are provided in tests/potentials. \\

\hl{
\textbf{Method Description:}

\textless Task Prompt \textgreater}
\end{promptquotebox}

\end{document}